\documentclass[prd,aps,showpacs,preprintnumbers,amssymb]{revtex4}
\usepackage{axodraw}
\usepackage{color}
\usepackage{epsf}

\def\e3p{$\eta \rightarrow 3 \pi$}

\begin{document}
\title{%
\hfill{\normalsize\vbox{%
\hbox{}
 }}\\
{About the role of scalars in a gauge theory}}

\author{Renata Jora
$^{\it \bf a}$~\footnote[2]{Email:
 rjora@theory.nipne.ro}}
\author{Salah Nasri
$^{\it \bf b}$~\footnote[3]{Email:
 nasri.salah@gmail.com}}

\affiliation{$^{\bf \it a}$ National Institute of Physics and Nuclear Engineering PO Box MG-6, Bucharest-Magurele, Romania}
\affiliation{$^{\bf \it b}$ Department of Physics, College of Science, United Arab Emirates University, Al-Ain, UAE}

\date{\today}

\begin{abstract}

We consider a new effective symmetry that acts on a gauge invariant Lagrangian. We show that the standard model after spontaneous symmetry breaking is invariant under this symmetry which identifies up to a scale factor  the gauge parameter with the elementary or composite scalars in the theory. We discuss some of the consequences of this symmetry in the abelian and non- abelian sectors of the standard model.
\end{abstract}
\maketitle

\section{Introduction}
The standard model of elementary particles \cite{Glashow}-\cite{Veltman} has become one of the most successful theories of modern physics with enumerable experimental confirmations culminating
with the discovery of a Higgs like particle at the LHC \cite{LHC}. However there are still theoretical issues to be clarified some of them in connection with the cosmological issues such as the origin of matter-antimatter asymmetry in the universe and the candidate for the dark matter. There are many established extensions of the standard model that solve some of these problems like the supersymmetric standard model, technicolor models, two Higgs doublet models, GUT models etc.
Among these only the experiment can decide.
The present work starts with two  fundamental question apparently unrelated: a) what is the role of the scalars in a gauge theory with fermions; b) why are the neutrino masses so small compare to the other fermions such that in first order approximation may be taken as zero? We will show here that a particular answer to these two problems  can be given in the context of a new symmetry of the  standard model Lagrangian.

\section{In the quest for a new symmetry}

Consider an abelian $U(1)$ theory with massless fermions:
\begin{eqnarray}
{\cal L}=\bar{\Psi}(i\gamma^{\mu}D_{\mu})\Psi-\frac{1}{4}(F_{\mu\nu})^2,
\label{lagr443}
\end{eqnarray}

where,
\begin{eqnarray}
D_{\mu}=\partial_{\mu}-i g A_{\mu}.
\label{cov554}
\end{eqnarray}

This Lagrangian not only preserves the Lorentz and gauge invariance but also the chirality of the states. It turns out that this theory is also invariant under the infinitesimal transformation given by the operator $K=\exp[k\gamma^{\mu}D_{\mu}]$ which is non-unitary and unbounded (and does not satisfy the premises of the Coleman Mandula theorem \cite{Coleman}):

 \begin{eqnarray}
&&\Psi'=\Psi+k\gamma^{\mu}D_{\mu}\Psi
\nonumber\\
&&A_{\mu}'=A_{\mu}
\label{transform}
\end{eqnarray}

 where k is an inverse scale such that for a square momentum $p^2$, $p^2k^2\ll 1$.
This is proven by:
\begin{eqnarray}
&&i\bar{\Psi'}\gamma^{\mu}D_{\mu}\Psi'=
i\bar{\Psi}\gamma^{\mu}D_{\mu}\psi+
\nonumber\\
&&i k(D_{\rho}\Psi)^{\dagger}\gamma^{\rho*}\gamma^0\gamma^{\mu}D_{\mu}\Psi+
i k \Psi^{\dagger}\gamma^0\gamma^{\mu}\gamma^{\rho}D_{\mu}D_{\rho}\Psi=
\nonumber\\
&&i\bar{\Psi}\gamma^{\mu}D_{\mu}\Psi
-ik\Psi^{\dagger}\gamma^0\gamma^{\rho}\gamma^{\mu}D_{\rho}D_{\mu}\Psi+
ik\Psi^{\dagger}\gamma^0\gamma^{\mu}\gamma^{\rho}D_{\mu}D_{\rho}\Psi=
\nonumber\\
&&i\bar{\Psi}\gamma^{\mu}D_{\mu}\Psi
\label{sym6657}
\end{eqnarray}

We generalize the Lagrangian in Eq. (\ref{lagr443}) to contain also a neutral scalar which couples with the fermion fields:
\begin{eqnarray}
{\cal L}=\frac{1}{2}(\partial_{\mu}B)^2-\frac{1}{2}m^2B^2+\bar{\Psi}(i\gamma^{\mu}D_{\mu})\Psi-m_f\bar{\Psi}\Psi-yB\bar{\Psi}\Psi-\frac{1}{4}(F_{\mu\nu})^2.
\label{lagr44343}
\end{eqnarray}

Here B is the scalar field and y is the Yukawa coupling for the fermion field. Applying  the infinitesimal transformation given by the operator K to the Yukawa term yields:
\begin{eqnarray}
-\Psi^{'\dagger}\gamma^0 B \Psi'&=&
-(K\Psi)^{\dagger}\gamma^0 B (K\Psi)
\nonumber\\
&=&-\Psi^{\dagger}\gamma^0 B\Psi-
k(D_{\mu}\Psi)^{\dagger}\gamma^{*\mu}\gamma^0B \Psi-k\Psi^{\dagger}\gamma^0\gamma^{\mu}BD_{\mu}\Psi
\nonumber\\
&=&-\Psi^{\dagger}\gamma^0 B\Psi+
k \Psi^{\dagger}\gamma^0\gamma^{\mu}BD_{\mu}\Psi-
k  \Psi^{\dagger}\gamma^0\gamma^{\mu}BD_{\mu}\Psi+
k \Psi^{\dagger}\gamma^0\gamma^{\mu}\partial_{\mu}B\Psi
\nonumber\\
&=&-\Psi^{\dagger}\gamma^0 B\Psi+k \Psi^{\dagger} \gamma^0\gamma^{\mu}\partial_{\mu}B\Psi.
\label{sym775665}
\end{eqnarray}

It turns out that this term is not invariant under the symmetry because of the extra term which contains a partial derivative of B. However the Lagrangian in Eq. (\ref{lagr44343}) is invariant under a slightly modified  symmetry given by the operator $K'=\exp[k\gamma^{\mu}D_{\mu}']$,  which for infinitesimal transformation  is defined as:
\begin{eqnarray}
\Psi'&=&\Psi+k\gamma^{\mu}D_{\mu}'\Psi
\nonumber\\
A_{\mu}' &=&A_{\mu}-\frac{k}{g}\partial_{\mu}B
\nonumber\\
B' &=& B
\label{sym6789}
\end{eqnarray}

where,
\begin{eqnarray}
D_{\mu}'=\partial_{\mu}-ig A_{\mu}'.
\label{rez4536}
\end{eqnarray}

The invariance of the Lagrangian (\ref{lagr44343}) under the symmetry defined  in (\ref{sym6789})  can be  proven as follows (for simplicity we set the Yukawa coupling $y=1$):
\begin{eqnarray}
{\cal L}'&=&
i\Psi^{\dagger}\gamma^0\gamma^{\mu}D_{\mu}'\Psi-\Psi^{\dagger}\gamma^0 B\Psi+
i k (\gamma^{\rho}D_{\rho}'\Psi)^{\dagger}\gamma^0\gamma^{\mu}D_{\mu}'\Psi
\nonumber\\
&&
+ik\Psi^{\dagger}\gamma^0\gamma^{\mu}D_{\mu}'\gamma^{\rho}D_{\rho}'\Psi -k (\gamma^{\rho}D_{\rho}'\Psi)^{\dagger}\gamma^0 B \Psi-
k \Psi^{\dagger}\gamma^0  \gamma^{\rho}D_{\rho}'\Psi
\nonumber\\
&=& i\Psi^{\dagger}\gamma^0\gamma^{\mu}D_{\mu}'\Psi+k\Psi^{\dagger}\gamma^0\gamma^{\mu}\partial_{\mu}B\Psi-\Psi^{\dagger}\gamma^0 B\Psi
\nonumber\\
&=&{\cal L}+g\Psi^{\dagger}\gamma^0\gamma^{\mu}(A_{\mu}'-A_{\mu})\Psi+k\Psi^{\dagger}\gamma^0\gamma^{\mu}\partial_{\mu}B\Psi.
\label{res553421}
\end{eqnarray}

where we have taken into account only the relevant terms. Note that the symmetry K' in (\ref{sym6789}) identifies the gauge parameter $\alpha$ up to a proportionality factor with the dynamical scalar in the model B, i.e. $\alpha=kB$.

For the  $K'$  to be a true symmetry of a gauge theory we need to prove that it is valid also for a non-abelian gauge theory. This is shown in,
\begin{eqnarray}
-\bar{\Psi}'B_1\Psi' &=&-\bar{\Psi}B_1\Psi -k(\gamma^{\rho}D_{\rho}'\Psi)^{\dagger}\gamma^0B_1\Psi-k\Psi^{\dagger}\gamma^0\gamma^{\rho}B_1D_{\rho}'\Psi=
\nonumber\\
&=&k\Psi^{\dagger}\gamma^0\gamma^{\rho}(D_{\rho}B_1)\Psi.
\label{rezult6578}
\end{eqnarray}

Here $B_1$ is a scalar field which may be elementary or composite and for the case of QCD  for example has the structure $B_1=B_1^a\frac{\lambda_a}{2}$. Thus a non-abelian gauge Lagrangian is invariant provided that,
\begin{eqnarray}
g(A_{\mu}'-A_{\mu})=-k D_{\mu}B_1,
\label{rez789}
\end{eqnarray}

So we  again obtain the relation $\alpha=kB_1$ but  this time for a non-abelian gauge parameter (for the case of QCD $\alpha=\alpha^a\frac{\lambda^a}{2}$). Hence we  showed  that the standard model of elementary particles is invariant under the infinitesimal symmetry given by the operator K' provided that his operator is adjusted to each fermion species and contains  the covariant derivative or simple derivatives corresponding to the each species kinetic term. For example the quarks transformations are given by:
\begin{eqnarray}
q'_i=q_i+k_i\gamma^{\mu}D_{\mu}q_i,
\label{cobder4546}
\end{eqnarray}

where i represents the quark flavor, $D_{\mu}=(\partial_{\mu}-ieA_{\mu}-igA^a_{\mu}\frac{\lambda_a}{2})$ is the full covariant derivative which contains the  electromagnetic  and the color gauge fields  $A_{\mu}$ and  $A_{\mu}^a$, respectively..

To complete the proof that  $K'$  is a  symmetry of the standard model Lagrangian we need to consider the transformation of the charged and neutral currents interaction. They have the form:
\begin{eqnarray}
{\cal L}_{CC} &=&\frac{g}{\sqrt{2}}(J_{\mu}^{+}W_{\mu}^{+}+h.c.)
\nonumber\\
{\cal L}_{NC}&=&\frac{g}{\cos\theta_W}J^0_{\mu}Z^{\mu}
\label{current678}
\end{eqnarray}

where the currents $J_{\mu}^{\pm,0}$ have  a structure of the type $\bar{f}^1_{L}\gamma^{\mu}\bar{f}^2_L$ for the charged currents or $\bar{f}^1_{L,R}\gamma^{\mu}\bar{f}^1_{L,R}$for the neutral ones.
We shall consider only the first case.  Under $K'$, a   left handed  fermion transforms as
\begin{eqnarray}
K'\Psi_L=\Psi_L+k\gamma^{\mu}D_{\mu}'(\frac{1-\gamma_5}{2})\Psi=\Psi_L+\frac{1+\gamma^5}{2}\gamma^{\mu}D_{\mu}'\Psi.
\label{rez443434}
\end{eqnarray}

Consequently, the action of this operator  on the charged current Lagrangian leads to,
\begin{eqnarray}
{\cal L}_{CC}'&=&\frac{g}{2}(K'f^1)^{\dagger}\gamma^0\gamma^{\mu}(K' f_2) W_{\mu}^+ +h.c.)
\frac{g}{2}(f^1)^{\dagger}\gamma^0\gamma^{\mu}f^2W_{\mu}^++
\nonumber\\
&=&k_1\frac{g}{2}(D_{\rho}' f^1)^{\dagger}\gamma^{\rho *}(\frac{1+\gamma^5}{2})\gamma^0\gamma^{\mu}(\frac{1-\gamma^5}{2})f^2W_{\mu}^++
k_2\frac{g}{2} f^{1\dagger}(\frac{1-\gamma^5}{2})\gamma^0\gamma^{\mu}(\frac{1+\gamma^5}{2})\gamma^{\rho}D_{\rho}'f^2W_{\mu}^++h.c.
\nonumber\\
&=&\frac{g}{2}(f^{1 \dagger}\gamma^0\gamma^{\mu}f^2W_{\mu}^++h.c.)={\cal L}_{CC}.
\label{res4435}
\end{eqnarray}

This concludes the proof that  the operator $K'$ , which contains all the unbroken gauge interactions specific to each fermion,  is a symmetry of   the standard model Lagrangian invariant .

\section{Connection between the new symmetry and confinement}

In section II we have shown that the symmetry given by the operator $K'=\exp[k\gamma^{\mu}D_{\mu}']$  works as an infinitesimal transformation acting on the fermion fields in  a gauge invariant Lagrangian. Thus we have made the underlying assumption that the constant $k=\frac{1}{\Lambda}$ functions as a natural cut-off of the theory such that all the momenta are lower than this  scale. We know that the gauge symmetry is a good symmetry both as infinitesimal transformation and finite one. It is natural to inquire what happens if the inverse scale k is of the size of the fermion momentum of one particular species of fermions.
Since we expect the nonabelian case to be more interesting we shall consider the QCD Lagrangian with only one flavor and with an additional quark-quark-scalar interaction,
 where the scalar may be elementary or composite:
\begin{eqnarray}
{\cal L}=-\frac{1}{2}{\rm Tr} G_{\mu\nu}G^{\mu\nu}+\bar{\Psi}(i\gamma^{\mu}D_{\mu}-m)\Psi-y\bar{\Psi}B_1\Psi.
\label{lagr5464}
\end{eqnarray}

We have proven in section II the invariance of this Lagrangian under the transformation:
\begin{eqnarray}
&&\Psi'=\Psi+k\gamma^{\mu}D_{\mu}'\Psi
\nonumber\\
&&A_{\mu}'=A_{\mu}-\frac{k}{g}D_{\mu}B_1
\nonumber\\
&&B_1'=B_1.
\label{rezult768}
\end{eqnarray}

We expect that the requirement that the Lagrangian in  Eq. (\ref{lagr5464}) be invariant under the full symmetry $K'$  will lead to particular solutions for the fields involved.
In what follows we  show that this is  indeed the case. So we require that
\begin{eqnarray}
(K'\Psi^{\dagger})\gamma^0(i\gamma^{\mu}D_{\mu}'-m-B_1)(K'\Psi)=\Psi^{\dagger}\gamma^0(i\gamma^{\mu}D_{\mu}-m-B_1)\Psi.
\label{eq332442}
\end{eqnarray}

The most obvious solution to the above equation  is then:
\begin{eqnarray}
\ln(K')\Psi &=&i\alpha\Psi
\nonumber\\
(-igk\gamma^{\mu}(A_{\mu}'-A_{\mu})+ikm+ikB_1)\Psi &=& i\alpha\Psi,
\label{rez4435}
\end{eqnarray}

where we used the equation of motion for the fermion field.

Now we reinforce    $kB_1=\alpha$, and $g(A_{\mu}'-A_{\mu})= - D_{\mu}\alpha$, to   obtain:
\begin{eqnarray}
k \gamma^{\mu}D_{\mu}B_1\Psi=m\Psi.
\label{equl8889}
\end{eqnarray}

Note that the equation in (\ref{equl8889}) is a constraint for both the fermion and the scalar fields. However we want to extract if possible only the solutions for
the scalar field.
First we apply the operators in Eq. (\ref{equl8889}) to the field $\bar{\Psi}$ and then multiply to the left to obtain:
\begin{eqnarray}
\bar{\Psi}k^2\gamma^{\mu}\gamma^{\nu}(D_{\mu}B_1)^{\dagger}(D_{\nu}B_1)\Psi=m^2\bar{\Psi}\Psi
\label{op7768}
\end{eqnarray}

Then we extract the trace condition with the append that the final solution we shall obtain from this satisfies also the full Eq. (\ref{op7768}):
\begin{eqnarray}
 {\rm Tr}[k^2\gamma^{\mu}\gamma^{\nu}(D_{\mu}B_1)^{\dagger}(D_{\nu}B_1)]=4m^2
 \label{rez5546}
 \end{eqnarray}

 This equation works as a constraint to the Klein Gordon equation in both cases where the field is elementary or as the equation of motion if the field is composite.
We simplify Eq. (\ref{rez5546}) to:
 \begin{eqnarray}
 \partial^{\mu}B_1^a\partial_{\mu}B_1^a+g^2f^{abc}A^{\mu b}B_1^cf^{amn}A_{\mu}^mB_1^n=\frac{m^2}{k^2}.
 \label{rez33245}
\end{eqnarray}

 One of the most interesting solutions to Eq. (\ref{rez33245}) is:
\begin{eqnarray}
 B_1=\frac{m}{k}(x_{\mu}-x_{0\mu})c^{\mu}\sum_a\frac{\lambda^a}{2},
 \label{sol8876}
 \end{eqnarray}

with $c_0=0$, $c_i=1$ and $A_{\mu}=0$.  Thus the field $B_1$ is an effective extended field with constant energy for which a suitable propagator is simply the function $\frac{m^2}{k^2}\delta(\vec{p})'$. An extra factor of $\delta(E-E_0)$ (where $E_0$ is the constant energy of the scalar field) which fixes the value of the energy should be introduced in the amplitudes. This insures the correct dimension of the corresponding propagator.

 We work in the nonrelativistic limit where the four momentum of a fermion is given by $(m,\vec{p})$ such that this type of solution make sense. Then considering the propagator in space time, it is obvious that the type of solution given in Eq. (\ref{sol8876}) applied between two pairs of fermion fields one placed at $x_0$ the other one at x gives a potential linearly growing with distance of confining type. This potential is attractive as it comes with a positive sign in the interaction hamiltonian.

 To see things more clearly let us estimate the potential for the scattering of four fermions intermediated by the field $B_1$. We can just simply replace in typical Yukawa calculations (see for example \cite{Peskin}) the propagator by $\frac{m^2}{k^2}\delta(\vec{p})'$ to obtain:

\begin{eqnarray}
V(q)=-i\frac{y^2m^2}{k^2}\delta(\vec{q})'
\end{eqnarray}
which in position space  gives
\begin{eqnarray}
V(r)=\int (-i\frac{y^2m^2}{k^2})\frac{d^3q}{(2\pi)^3}e^{i\vec{q}\vec{x}}\delta(\vec{q})'=-\frac{y^2m^2}{k^2}r
\label{rez3245}
\end{eqnarray}

 Note that analogously to the Wilson loop, the field B correspond to the gauge parameter of the non abelian gauge theory but the confining behavior is obtained quite straightforwardly by simply reinforcing the gauge symmetry.

\section{Naturalness in the context of the new symmetry}

After spontaneous symmetry breakdown of the electroweak group the standard model of elementary particles possesses a $U(1)\times SU(3)$ symmetry. Let us consider only the U(1) symmetry.
According to the picture depicted in section II the invariance of the Lagrangian under the symmetry $K'$ in Eq. (\ref{sym6789}) requires  a neutral scalar B that couples with the fermions
such that the relation,
\begin{eqnarray}
kB=\alpha
\label{rel887}
\end{eqnarray}

holds where k is a parameter with dimension ${\rm mass}^{-1}$ related to the cut-off scale of the theory. Since the parameter $\alpha$ is universal for all the fermions the inverse scale $k_f$ specific to one fermion flavor should be adjusted such that $y_fk_f/q_f=k$ where $q_f$ is the charge and $y_f$ is the corresponding Yukawa coupling.

However if we associate the gauge parameter with a dynamical scalar we have a problem with fixing the gauge in the usual manner. Note that up to this point there is nothing to prevent
the scalar B from having a mass and a potential and to behave like a regular Higgs boson.

Let us Consider a U(1) gauge theory with a neutral scalar and fermions. The generating functional is given by:
\begin{eqnarray}
Z[B,A_{\mu},\Psi]=\int {\cal D}\bar{\Psi}{\cal D}\Psi{\cal D}A_{\mu}{\cal D}B\exp[i\int d^4x {\cal L}].
\label{genfunc67}
\end{eqnarray}\

In the standard approach  the gauge fixing condition is introduced through:
\begin{eqnarray}
\int {\cal D} A\exp[i S[A]]=\det(\frac{\delta(G(A^{\alpha}))}{\delta\alpha}\int {\cal D}\alpha\int {\cal D}A\exp[i S[A]]\delta(G(A)),
\label{fix453}
\end{eqnarray}

where,
\begin{eqnarray}
G(A^{\alpha})=\partial^{\mu}A_{\mu}-\frac{1}{g}\partial^{\mu}\partial_{\mu}\alpha.
\label{g778}
\end{eqnarray}

For our case however the condition given in Eq. (\ref{g778}) may seem like a good gauge fixing condition but it is not since it is equivalent to:
\begin{eqnarray}
\partial^{\mu}A_{\mu}'=\partial^{\mu}A_{\mu}-\frac{k}{g}\partial^{\mu}\partial_{\mu}B
\label{byyu}
\end{eqnarray}

and the field B is dynamical and has its own equation of motion. We will show however that one can apply a similar system of constraints to the case at hand. We start by fixing the gauge field for $\xi=1$, the Feynman gauge:

\begin{eqnarray}
\int {\cal D}A\exp[i S[A]] &=&{\rm const} \int {\cal D}A\exp[i S[A]]\delta(\partial^{\mu}A_{\mu}-\omega(x))
\nonumber\\
&=&{\rm const}'\int {\cal D}\omega\int {\cal D} A \exp[i S[A]]\delta(\partial^{\mu}A_{\mu}-\omega(x))\exp[-i\int d^4x\frac{\omega^2}{2}]
\nonumber\\
&=&{\rm const}'\exp[{i S[A]}] \exp[-i\int d^4x\frac{1}{2}(\partial^{\mu}A_{\mu})^2].
\label{gf556}
\end{eqnarray}

We shall use a trick to obtain the correct gauge condition in this set-up. We know that,
\begin{eqnarray}
\omega(x)=-\frac{k}{g}\partial^{\mu}\partial_{\mu}B=\frac{k}{g}m_0^2B
\label{mot667}
\end{eqnarray}

where we used the Klein Gordon equation for the scalar field. Then we need a term,
\begin{eqnarray}
\exp[-i\int d^4x \frac{k^2m_0^4}{g^2}B^2]
\label{term7768}
\end{eqnarray}

in the Lagrangian. This should come from a simple Lagrangian for the scalar field:
\begin{eqnarray}
{\cal L}=\frac{1}{2}(\partial_{\mu}B)^2-\frac{1}{2}m_0B^2
\label{lagr}
\end{eqnarray}

We thus require,
\begin{eqnarray}
&&\int d^4 x[\frac{1}{2}(\partial_{\mu}B)^2-\frac{1}{2}m_0B^2]=\int d^4x (-\frac{k^2m_0^4B^2}{g^2})
\nonumber\\
&&\partial^{\mu}\partial_{\mu}B-(m_0^2-\frac{k^2m_0^4}{g^2})B=0
\label{rez556}
\end{eqnarray}

in all orders of perturbation theory. This is equivalent to asking for the following relation between the physical and bare mass:
\begin{eqnarray}
m^2=m_0^2(1-\frac{k^2m_0^2}{g^2})
\label{ren6657}
\end{eqnarray}

where the sign of the correction can also be positive.
Practically the condition (\ref{ren6657}) fixes the physical mass of the Higgs boson in terms of the cut-off scale of the theory $\Lambda=\frac{1}{k}$.  Alternatively one can compute corrections to the Higgs mass and determine the inverse scale k.

In summary the fixed generating functional for the gauge Lagrangian will become:

\begin{eqnarray}
Z[\bar{\Psi},\Psi,A,B]=
{\rm const} \int {\cal D}\bar{\Psi}{\cal D}\Psi{\cal D}A{\cal D}B\exp[i\int d^4x {\cal L}]\exp[-i\int d^4x \frac{\omega^2}{2}]\delta(\partial^{\mu}\partial_{\mu}B+m^2B)
\delta(\partial^{\mu}A_{\mu}-\omega)
\label{rez55467}
\end{eqnarray}

Note that we have the usual gauge fixing as it should be with an extra constraint  for the scalar mass which fixes its behavior for some type of Lagrangians with the cutoff scale.
For the initial symmetry K to work however one needs a cut-off scale much larger than all the masses in the theory.  The larger the scale the better all physical quantities behave.  The constraints on the scalar particle are milder in the presence of multiple scalars which couple with different flavors of fermions.

\section{Conclusions and discussion}

Some of the issues associated with the standard model after spontaneous symmetry breaking can be solved if one considers an effective symmetry (the $K'$ symmetry) acting on the fermions in the Lagrangian.
This symmetry has some similarities with the BRST symmetry \cite{BRST} but also some important differences. First of all  it is an effective symmetry that depends on the natural cut-off scale in the theory;
secondly it identifies the gauge parameters with the scalars, elementary or composite that exist in the theory. However as opposed to the BRST symmetry the scalars are dynamical and present in the theory and one of them can correspond to the Higgs boson found at the LHC. Moreover this simple symmetry can explain why neutrinos which do not have any vector gauge interactions are not coupled at tree level with any scalars ( e.g.   an $ SU(2)_L$  scalar  triplet for  the case of Majorana neutrinos ) :  such a  coupling would break the K symmetry.

In any gauge theory one needs  a gauge condition necessary to eliminate the redundant degrees of freedom. We did that (for the U(1)  gauge group) in the context of the K symmetry by fixing regularly the gauge field condition which produces an extra constraint related to the scalar field. Thus  its mass is fixed by the theory and the off-shell degrees of freedom are eliminated. This might create some problems with unitarity  of the model for the case of a single scalar.  However this new symmetry is better realized in the presence of multiple scalars, case in which some of the scalars survive off-shell.

We  have also  showed that the scalar field which is proportional to the gauge parameter for the group $SU(3)_C$ can play the role of the Wilson loop for describing confining. Note that this scalar may be elementary which contradicts the observation, or composite which better fits the general picture of QCD. In the latter case the scalar is introduced in theory as an auxiliary field which is eliminated by the equation of motion, thus leading to four fermion interactions.

 We shall postpone discussions of the implications of this symmetry for extensions of the standard  model in general for further work.

\section*{Acknowledgments} \vskip -.5cm
The work of R. J. was supported by a grant of the Ministry of National Education, CNCS-UEFISCDI, project number PN-II-ID-PCE-2012-4-0078.

\end{document}